\documentclass[aps,showpacs,preprint,nofootinbib]{revtex4-1}
\usepackage{graphicx,color}
\usepackage[colorlinks]{hyperref}
\pdfoutput=1
\usepackage{calc}
\usepackage{subfig}

\usepackage[T1]{fontenc}
\usepackage{array}
\usepackage{makecell}
\newcolumntype{x}[1]{>{\centering\arraybackslash}p{#1}}

\usepackage{tikz}
\newcommand\diag[4]{%
  \multicolumn{1}{|p{#2}|}{\hskip-\tabcolsep
  $\vcenter{\begin{tikzpicture}[baseline=0,anchor=south west,inner sep=#1]
  \path[use as bounding box] (0,0) rectangle (#2+2\tabcolsep,\baselineskip);
  \node[minimum width={#2+2\tabcolsep},minimum height=\baselineskip+\extrarowheight] (box) {};
  \draw (box.north west) -- (box.south east);
  \node[anchor=south west] at (box.south west) {#3};
  \node[anchor=north east] at (box.north east) {#4};
 \end{tikzpicture}}$\hskip-\tabcolsep}}
\begin{document}
\def\ben{\begin{eqnarray}}
\def\en{\end{eqnarray}}
\def\non{\nonumber}
\def\la{\langle}
\def\ra{\rangle}

\def\mlstar{m_{\scriptstyle{l^*}}}
\def\mzprime{m_{\scriptstyle{Z'}}}
\def\mz{m_{\scriptstyle{Z}}}

\title{Production and decay rates of excited leptons 
in a left-right symmetric scenario}

\author{Piyali Banerjee}
\email{banerjee.piyali3@gmail.com} 
\author{Urjit A. Yajnik}
\email{yajnik@iitb.ac.in} 

\affiliation{ 
Department of Physics, 
Indian Institute of Technology Bombay, 
Mumbai 400076, 
India
}

\begin{abstract}

We merge two  leading Beyond Standard Model scenarios, 
namely compositeness and  left-right symmetry, and probe the resulting 
collider signatures in the  leptonic case. 
The constraints on composite models for fermions leave open the possibility
of vector like excitations of Standard Model (SM) fermions. Here we consider
the possibility of  low scale left-right gauge symmetry 
$SU(2)_{R} \times SU(2)_{L} \times U(1)_ {B-L}$, with  the simplifying assumption 
that the right like excited sector of fermions is  significantly heavier than 
the excitations of the left chiral fermions. It is found that the right handed 
currents still contribute to observable processes, and alter
the existing bounds on the scale of compositeness. 
The cross section times branching  ratio of the photon decay channel is 
strongly depressed, bringing down  the exclusion limit of the mass of excited electrons 
from about 2 TeV to below 1 TeV. On the other 
hand, cross section times branching ratio of the Z decay channel is  significantly 
enhanced and remains
greater than that of the photon  channel. We thus propose analyzing the 
Z decay channel in existing 
collider data in order to search for signature of left-right symmetry as 
well as excited leptons with masses above 1 TeV. 

\end{abstract}

\maketitle 

\section{Introduction} 
One way around the conceptual problems faced by the Higgs mechanism operating at scales 
much lower than the Planck scale, is to conjecture, as has been done in many works, 
that the observed spectrum of particles arises as a set of composites 
of some more fundamental particles. If this is true then in particular, there would be
excited states heavier than known fermions, with some correspondence to, and
decay channels into, known particles.
Such models have been considered early in the development of
electroweak theory, \cite{Terazawa:1977, Neeman:1979, Barbieri:1981cy,Terazawa:1982,
Harari:1979gi,Harari:1980ez,Shupe:1979fv} from theoretical motivations,
and also subsequently, from a phenomenological point of view in response 
to emerging experimental signatures \cite{Renard:1982ij,Renard:1983ks,
Cabibbo:1983bk,Eichten:1983hw,Hagiwara:1985wt,Rujula:1984, Kuhn:1984}.
Signatures of excited leptons have been searched for at the DESY 
Hadron Electron Ring Accelerator (HERA) and the CERN Large Electron 
Positron collider (LEP) \cite{hagiwara:1985, Boudjema:1993}, the 
Fermilab Tevatron collider \cite{Acosta:2005, Abazov:2008}, 
the CERN Large Hadron Collider (LHC) \cite{Baur:1990,Cakir:2004,Biondini:2012ny}, 
and are proposed for the next linear colliders \cite{Cakir:2004lin, Cakir:2008}. Up to now, 
no signal has been observed for excited states \cite{Aad:2013jja}, but with 
higher centre-of-mass energy and  more data we would be able make the search more robust. 

The confirmation of small neutrino masses in the past decade 
\cite{Fukuda:2001nk,Ahmad:2002jz,Ahmad:2002ka,Bahcall:2004mz} has raised 
the possibility that the spectrum of matter is after all symmetric 
between the two handedness states, and the  maximal parity 
violation of Weak forces is only a low energy effect. Chirality 
already incorporated in the Standard Model (SM) however continues to be 
the underlying principle on which building blocks of matter are 
organized, with the advantage of universal source of masses from 
spontaneous symmetry breaking. While the question of grand unification 
seems to have got postponed with the absence of proton decay at expected 
energies, we may take the Left-Right symmetric model (LRSM) as a natural 
extension of the SM. The Left-Right symmetric model 
\cite{Pati:1974yy,Mohapatra:1974gc} treats both left- and right-handed 
fermions as doublets, and possesses the potential to provide an elegant 
explanation of neutrino masses \cite{Mohapatra:1979ia} through the 
see-saw mechanism 
\cite{Minkowski:1977sc,GellMann:1980vs,Yanagida:1979as}. Additionally 
the model gauges the $B-L$ quantum number~\cite{Mohapatra:1980qe}, the 
only anomaly free quantum number of the the SM left ungauged. 

The scale of parity 
breakdown is as yet unknown. The see-saw mechanism \cite{GellMann:1980vs,Yanagida:1979as}, while 
providing an elegant qualitative explanation, is unable to predict the 
parity breaking energy scale due to wide variation in the fermion masses 
across the generations. The scale of the parity breakdown, equivalently that of the
right handed Majorana neutrinos is usually pegged high, $10^{14}$GeV. However
this is subject to the choice of the "pivot" mass, which is often pegged at the electroweak
scale. If the pivot mass is that of the lighter charged fermions, it is equally well
possible to have the appearance of the right handed gauge forces at as low as 
TeV scale \cite{Sahu:2004sb}.
The current experimental constraints on $Z'$ mass are as low as the TeV scale \cite{Beringer:1900zz}. 
However the existing models of excited fermions, based on the SM, have 
not attempted to address the incorporation of non-zero neutrino masses. 
So it is natural to attempt combining  these two models to address the above 
inadequacies of SM. We thus have 
two new scales appearing beyond the SM, the scale $M_R$ which determines 
the mass of $W_R$ and $Z_R$ bosons, and hence the mass $\mzprime$ of 
new neutral gauge boson $Z'$, and separately, a scale $\Lambda$, which 
determines the mass $\mlstar$ of the composite fermions.

Any model of excited fermions is strongly constrained, \cite{Brodsky:1980zm}
\cite{Renard:1982ij} firstly by the 
absence of any structure for the fermions upto currently accessible 
energies, and secondly by the anomalous
magnetic moments $g-2$ as well the electric dipole moments of the leptons. 
The contribution of vector like fermionic states to $g-2$ is however
significantly suppressed and therefore such excited states remain light
enough to be accessible to current and near future accelerators without
contradicting  the other data.

Incorporating Left-Right symmetry in this setting 
requires doubling up the entire spectrum of the excited fermions, one 
set connected to the left chiral fermions and another connected to the 
right chiral fermions. To keep the discussion simple in this initial 
investigation we assume the heavy vector like fermions to be doublets 
only under the SM $SU(2)_L$, and ignore the spectrum that may be 
associated to the $SU(2)_R$ of the LRSM. This may actually be a valid 
approximation in the full fledged model if the $SU(2)_R$ related vector 
like heavy spectrum is naturally heavier than the left like. This 
circumstance could have the same underlying physics as that which 
suppresses the right handed currents at low energies.  With above 
reasoning, in this paper we work in the simplified framework where the 
excited fermion 
spectrum to be explored is coupled directly only to the SM 
$SU(2)_L$. Nevertheless, the $l^*$ sector leaves its stamp on the search 
strategies for Left-Right symmetry. This is because in LRSM, the 
observed $Z$ boson is an admixture of $Z_L$ and $Z_R$ and the signal for 
$Z'$ bosons would have to be modified if this excited lepton sector 
exists. In the following we shall assume 
the hierarchy $\Lambda$ $\gtrsim \mlstar >$ $\mzprime > \mz$.
 
This paper is organised as follows. In Sec.~\ref{sec:setup} we introduce 
the effective Lagrangian describing the interaction of excited leptons with left right symmetric fermions. 
We then analyze the various production and decay rates 
of $l^{*}$ for different $Z^{'}$ and $W^{'}$ mass, 
and the last section contains a summary and conclusions.
 
%%%%%%%%%%%%%%%%%%%%%%%%
\section{Extension to left-right symmetric case}
\label{sec:setup}
%%%%%%%%%%%%%%%%%%%%%%%%

The $SU(2)_{L} \times U(1)_{Y}$ invariant effective Lagrangian  that describes the
interaction between an ordinary lepton $l$, a  gauge boson V(=W, Z, $\gamma$), and an
excited lepton $l^{*}$ is introduced as follows \cite{Cabibbo:1983bk}\cite{Hagiwara:1985wt}\cite{Baur:1990}: 
\begin{equation}
\label{eq:SMLag}
\mathcal{L}_{\mbox{trans}} =
\frac{1}{2\Lambda} \bar{l}^*_R \sigma^{\mu \nu}
\left[
g_s f_s \frac{\lambda^a}{2} G^a_{\mu \nu} +
g f \frac{\mathbf{\tau}}{2} \cdot \mathbf{W}_{\mu \nu} +
g' f' \frac{Y}{2} B_{\mu \nu}
\right]
l_L +
\mbox{H.c.}
\end{equation}
in a notation that parallels \cite{Baur:1990}. Here $\Lambda$ is the compositeness scale, 
and the $G$, $W$ and $B$ denote the field strengths of the colour and electroweak sectors.
$f$ and  $f_0$ are the transition magnetic moments 
arising from the compositeness dynamics.
Further, an underlying assumption of excited fermions
hypothesis is that no mixing of generations is triggered by the 
excitation physics. In this paper we shall refers to
this as the Standard Model based or SM based approach, and compare specific 
calculations to the results of \cite{Baur:1990}, referred to as  BSZ. 

Left-right symmetric extension of the SM ensures parity symmetry while retaining the highly desirable
chiral nature of the fermionic spectrum.  The extension entails that we add a right handed
neutrino state to each generation of the spectrum, and together with this, the SM fermions have additional 
quantum numbers under an extended gauge group 
$SU(3)_c\otimes SU(2)_L\otimes SU(2)_R\otimes U(1)_{B-L}$ 
which displayed for one generation are
\begin{equation}
l_L=
\left(\begin{array}{c}
\ \nu_L \\
\ e_L
\end{array}\right)
\sim (1,2,1,-1)\hspace*{0.8cm}
l_R=
\left(\begin{array}{c}
\ \nu_R \\
\ e_R
\end{array}\right)
\sim (1,1,2,-1)
\end{equation}
Similar changes to the quarks sector are required, but not relevant to this study.
The corresponding electric charge formula becomes left-right symmetric, $Q_{\rm electric}=$ 
$T^3_L+T^3_R+\frac{1}{2}(B-L)$. To this list of particles, we add, for each generation, 
excited leptons, being a pair of doublets, taken together forming a vector like 
representation of $SU(2)_L$ 
\begin{equation}
l^*_L=
\left(\begin{array}{c}
\ \nu^*_L \\
\ e^*_L
\end{array}\right)
%\hspace*{0.3cm}
\sim
%\hspace*{0.3cm}
%\sim (1,2,1,-1),
%\hspace*{0.3cm}
l^*_R=
\left(\begin{array}{c}
\ \nu^*_R \\
\ e^*_R
\end{array}\right)
\sim (1,2,1,-1)
\end{equation}
There should also exist a similar set of vector like excited leptons with charges $(2,1,1,-1)$.
As explained in the Introduction, we postpone this study for the future, assuming for the purpose
of this paper that the $SU(2)_R$ excited fermions are too heavy to affect experimental signatures.

We therefore propose the generalisation of the effective Lagrangian of Eq. (\ref{eq:SMLag}) 
to the left-right symmetric case in the form
\begin{equation}
\label{eq:LRLag}
\mathcal{L}_{\mbox{trans}} =
\frac{1}{2\Lambda} \bar{l}^*_L \sigma^{\mu \nu}
\left[
g_s f_s \frac{\lambda^a}{2} G^a_{\mu \nu} +
g_{1} f_{1} \frac{\tau}{2} \cdot \mathbf{W}^{L}_{\mu \nu} +
g_{2} f_{2} \frac{\tau}{2} \cdot \mathbf{W}^{R}_{\mu \nu} +
g'' f'' \frac{B-L}{2} B^{B-L}_{\mu \nu}
\right]
l_R +
\mbox{H.c.}
\end{equation}
Here $W^{L}_{\mu\nu}$ and $W^{R}_{\mu\nu}$ are the field strength 
tensors of $SU(2)_{L}$ and $SU(2)_{R}$ gauge fields
respectively and $B^{B-L}_{\mu\nu}$
is the field strength tensor of $U(1)_{B-L}$. 
$g_{1}$, $g_{2}$ and $g''$
are the $SU(2)_L$, the $SU(2)_R$ and the $U(1)_{B-L}$ gauge couplings respectively. 
$f_{1}$,  $f_{2}$ and $f^{''}$ are the new couplings that arise due  to
compositeness in the theory. 
Consistent with the original left-right symmetry philosophy we assume 
$g_{1}= g_{2}$, in turn equal to $g$ of Eq. (\ref{eq:SMLag}). 

However the effective couplings $f_1$, $f_2$, $f′′$ above are constrained by 
their potential contribution to the anomalous magnetic moment form 
factors of the known leptons at one-loop
level. One approach to calculating the loop contribution is to introduce a 
dipolar form factor for the excited leptons, given by 
$\frac{\Lambda^4}{(q^2 - \Lambda^2)^2}$ \cite{Renard:1982,Rakshit:2001}, 
where $q^2$ is the virtual photon mass squared.
%The contributions of the other one-loop
%diagrams, though individually non-zero, cancel amongst each other.
Let $\frac{g-2}{2}$ denote the anomalous magnetic moment of the muon.
Let  $m$, $M$ denote the masses of ordinary and excited muons respectively, and
$\Lambda$ the compositeness scale. Under the parameter regime $m \ll M \leq \Lambda$,
$M$ of the order $\Lambda$,
our calculation gives us a constraint on $f_{1}$,  $f_{2}$ and $f^{''}$ 
described by the equation 
\begin{equation}
\label{eq:magmoment}
(f_1 + f_2 + f'')^2 
\frac{m M}{\Lambda^2} \cdot I(M^2/\Lambda^2)
\leq \frac{(g-2)\pi}{\alpha}
\end{equation}
Above, $I$ is a function of $M^2/\Lambda^2$ only. When $M$ is of the order of 
$\Lambda$, $I$ is of the order of $10^{-1}$. In particular, when $M = \Lambda$,
$I(1) = 7/45$. Since we consider $\Lambda$ at the TeV scale, $m/M$ is of the
order $10^{-4}$. Using the best value of $g-2$ and $\alpha$ known currently
\cite{Davier-Marciano:2004}, the right hand side of the above expression turns
out to be $2.066 \times 10^{-6}$. This constrains $f_1 + f_2 + f''$ to be 
around $0.5$ for our range of parameters.

The results of the GERDA experiment \cite{Agostini:2013} have set a 
limit of $3 \times 10^{25}$ years on the half life of neutrinoless 
double beta decay of ${}^{76}\mbox{Ge}$. Neutrinoless double beta 
decay is another lepton flavour violating process, which constrains 
$f_1$, $f_2$ and $\Lambda$ in our model. Using the above value for the 
half life and working on the lines of \cite{Panella:1995}, we get a 
constraint of $f_1 < 1.68$ for $\Lambda = 1~\mbox{TeV}$ and excited 
neutrino mass $600~\mbox{GeV}$, and a loose constraint of 
$\Lambda > 118~\mbox{GeV}$ for $f_1 = 0.2$ and excited neutrino mass 
$600~\mbox{GeV}$. The same constraints hold between $f_2$ and $\Lambda$ 
as we assume $f_1 = f_2$, that is, the strengths of the couplings to 
left and right handed fermions are the same. The constraints on 
$f_1$ and $\Lambda$ are looser for higher values of the excited neutrino 
mass, as can be seen from Table~\ref{table:doublebetadecay}.
\begin{table}
\setlength{\extrarowheight}{0.3cm}
\begin{tabular}{|x{3.2cm}|x{1.3cm}|x{1.3cm}|x{1.3cm}|x{1.3cm}|x{1.3cm}|}
\hline
\diag{.1em}{3.2cm}{$m_{\mu^*}~\mbox{(GeV)}$}{$\Lambda~\mbox{(GeV)}$}
& 1000 & 2000 & 3000 & 4000 & 5000 \\
\hline
600
& 1.68 & 3.37 & 5.05 & 6.74 & 8.42 \\
\hline
800
& 1.94 & 3.89 & 5.83 & 7.78 & 9.72 \\
\hline
1000
& 2.17 & 4.35 & 6.52 & 8.70 & 10.87 \\
\hline
1600
& 2.75 & 5.50 & 8.25 & 11.00 & 13.75 \\
\hline
2000
& 3.08 & 6.15 & 9.23 & 12.30 & 15.38 \\
\hline
\end{tabular}

\medskip

\begin{tabular}{|x{3.2cm}|x{1.3cm}|x{1.3cm}|x{1.3cm}|x{1.3cm}|x{1.3cm}|}
\hline
\diag{.1em}{3.2cm}{$m_{\mu^*}~\mbox{(GeV)}$}{$f_1$}
& 0.1 & 0.2 & 0.5 & 1.0 & 2.0 \\
\hline
600
& 59.37 & 118.74 & 296.86 & 593.71 & 1184.42 \\
\hline
800
& 51.42 & 102.83 & 257.09 & 514.17 & 1028.34 \\
\hline
1000
& 45.99 & 91.98 & 229.94 & 459.89 & 919.77 \\
\hline
1600
& 36.36 & 72.71 & 181.79 & 363.57 & 727.15 \\
\hline
2000
& 32.52 & 65.04 & 162.60 & 325.19 & 650.38 \\
\hline
\end{tabular}

\caption{
Constraints on $f_1$ and $\Lambda$ for different values of the 
excited neutrino mass arising from the GERDA neutrinoless double beta 
decay experiment. Top: Upper bounds on $f_1$ 
Bottom: Lower bounds on $\Lambda~\mbox{(GeV)}$.
}
\label{table:doublebetadecay}
\end{table}

We also consider the possibility of whether the $g − 2$ measurements can 
constrain flavour changing transitions like $\mu \rightarrow e \gamma$ in 
our model. Arguing along the lines of \cite{Renard:1982}, and using the 
best value of $1.7 \times 10^{-32}~\mbox{GeV}$ for the partial width 
$\Gamma_{\mu \rightarrow e \gamma}$ known currently from the 
MEG experiment \cite{meg:2013}, we get that the constraint will be 
ineffectual unless $\Lambda > 10^{11}~\mbox{GeV}$.

The partial widths for the electroweak decay channels \cite{Baur:1990} are now  given,
neglecting ordinary quark masses, by the formulae
\begin{equation}
\begin{array}{r c l}
\Gamma(f^* \rightarrow f \gamma) 
& = &
\frac{1}{4} \alpha C_\gamma^2 \frac{{m^*}^3}{\Lambda^2}, \\
\Gamma(f^* \rightarrow f V) 
& = &
\frac{1}{8} \frac{g_V^2}{4 \pi} C_V^2 \frac{{m^*}^3}{\Lambda^2}
\left(1 - \frac{m_V^2}{{m^*}^2}\right)
\left(2 + \frac{m_V^2}{{m^*}^2}\right).
\end{array}
\end{equation}

where ($V = W, W', Z, Z'$) and
\begin{equation}
\begin{array}{l l}
C_{\gamma} =  
 f_{1}T_{3} + f_{2} T_{3} + f'' \frac{B-L}{2}; 
& \\
C_{Z} =
f_{1} T_{3} \cos^2 \theta_W - f_{2} T_{3} \sin^2 \theta_W + 
f'' \frac{B-L}{2};~~~
&
C_{Z'} = 
f_{2} T_{3} \cot \theta_R + f" \tan \theta_R \frac{B-L}{2}; \\
C_{W} = \frac{f_{1}}{\sqrt{2}}; 
&
C_{W'} = \frac{f_{2}}{\sqrt{2}}. 
\end{array}
\end{equation}
where $\theta_{W}$ is the weak mixing angle and 
\begin{equation}
\sin^2 \theta_{R} = 
\frac{2 \sin^2 \theta_{W}}{1 + \sin^2 \theta_{W}}.
\end{equation}
is the angle mixing $T_{3R}$ and $B-L$ giving rise to right handed neutral currents.

%\section{Effects of new decay channels}
%\label{sec:numerical}

We study the decay rates of the excited leptons in the context of above interaction,
and try to identify how the signals get  modified compared to the SM based approach.
In the SM based approach, the excited lepton has available to it the decay channels 
$l,\gamma$; $l,Z$; $\nu_{l},W$. 
Extension to the left-right group leads to the possibility of 
two additional decay modes namely
$l,Z'$; $\nu_{l},W'$. We study the effect on the production cross
section of $\mu^{*}$ of these additional two channels as a function of
$\mu^{*}$ mass $m_{\mu^{*}}$ and compositeness scale $\Lambda$. We also vary the $Z'$ mass to
see the effect on production cross-section times branching ratios (B.R.) of $\mu^{*}$.

To carry out the main comparison, we have studied the case of $\mu^{*}$. The contributions 
from $e^{*}$ would be similar in principle, except for the difference in the input mass of the 
usual lepton. Since  the effect of the  mass of the usual leptons being negligible at the scale 
of the scattering  we get similar results for $e^{*}$ and $\mu^{*}$. 

For the simulation of the signal a customized
 version of the Pythia8 event generator \cite{Sjostrand:2007gs} is used, 
following the model of \cite{Baur:1990}. Decays via contact interactions, 
not implemented in Pythia8, contribute between a few percent 
of all decays for $\Lambda \gg m_{\mu^{*}}$ and 92\% for $\Lambda =
m_{\mu^{*}}$  \cite{Baur:1990,Cakir:2004}. However, in this study 
 we are only interested in the new production cross section and the
 branching ratios (B.R.) as the outcome of the Lagrangian given by Eq. (\ref{eq:LRLag}). 
 So for all purpose we choose to neglect the correction for now.    

\section{Results and Discussions}
\label{sec:results}

The total production cross section versus $m_{\mu^{*}}$ for both SM based approach
of \cite{Baur:1990} and for our LRSM based approach are shown for $\mu^{*}$ in Fig.~\ref{muxsec}.
We also show the variation of branching ratios for
the different gauge decay modes as a function of $m_{\mu^{*}}$  
in both the appraoches in Fig.~\ref{mubr}.
We then compare the cross section times branching ratios for the photon and $Z$ channels 
as a function of $m_{\mu^{*}}$ 
in both the approaches in Fig.~\ref{muxsectimesbr}. In all these figures, 
the coupling parameters $f_1$, $f_2$, $f''$ are set to be equal under the constraint
of Equation~\ref{eq:magmoment}.

\newlength{\figwidth}
\setlength{\figwidth}{\textwidth/2}
\begin{figure}
\begin{tabular}{c c}
\subfloat[]{\includegraphics[width=\figwidth]{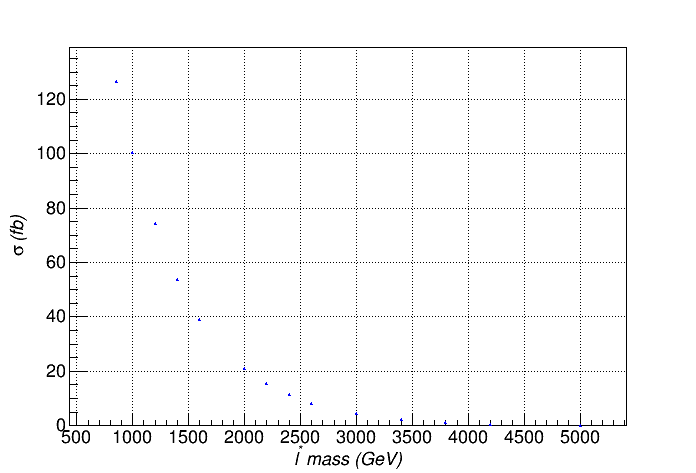}}
&
\subfloat[]{\includegraphics[width=\figwidth]{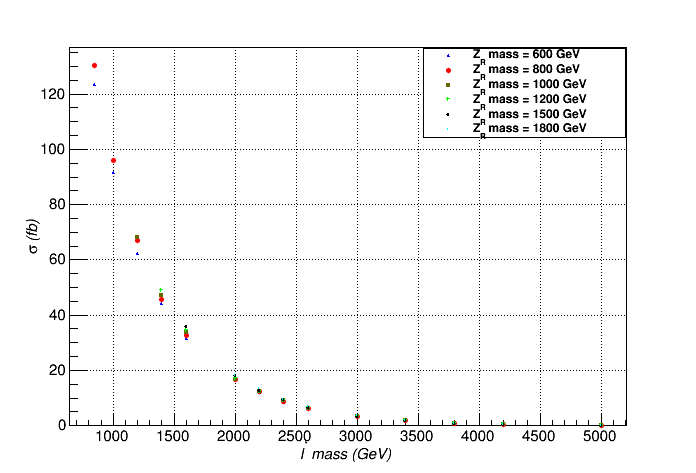}}
\\
\end{tabular}
\caption{
Cross section as function of $\mu^*$ mass, $\Lambda = 5000$ GeV, $14$ TeV $pp$ beam.
Left:  BSZ, as in Ref. \cite{Baur:1990}.
Right: Left-right symmetric theory for $W^{'}$ mass 800 GeV and
       six different $Z'$ masses, $f_1 = f_2 = f''$.}
\label{muxsec}
\end{figure}

\begin{figure}
\begin{tabular}{c c}
\subfloat[]{\includegraphics[width=\figwidth]{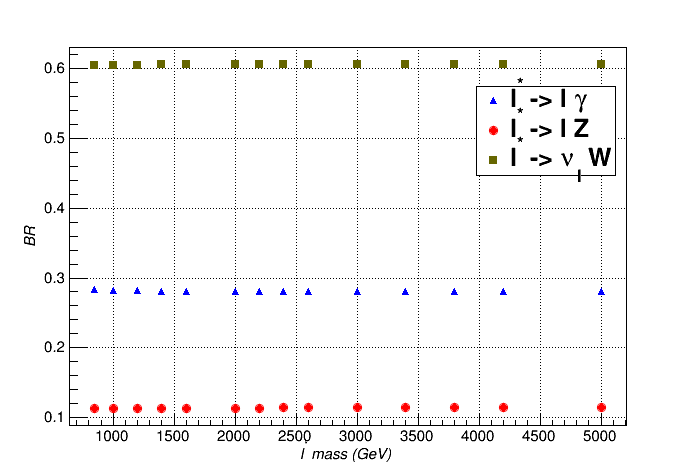}}
&
\subfloat[]{\includegraphics[width=\figwidth]{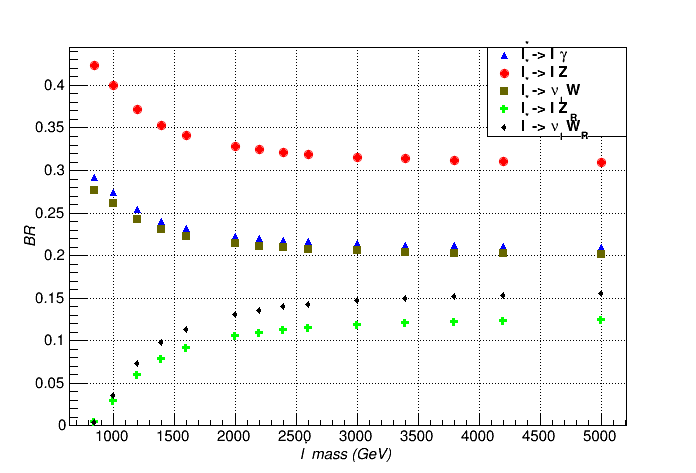}}
\\
\subfloat[]{\includegraphics[width=\figwidth]{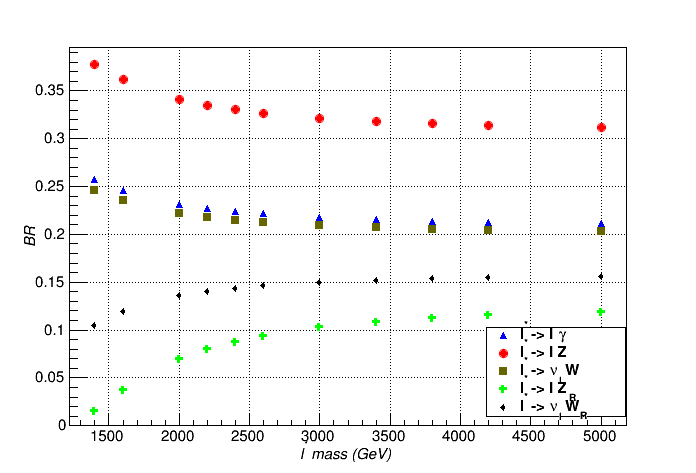}}
&
\subfloat[]{\includegraphics[width=\figwidth]{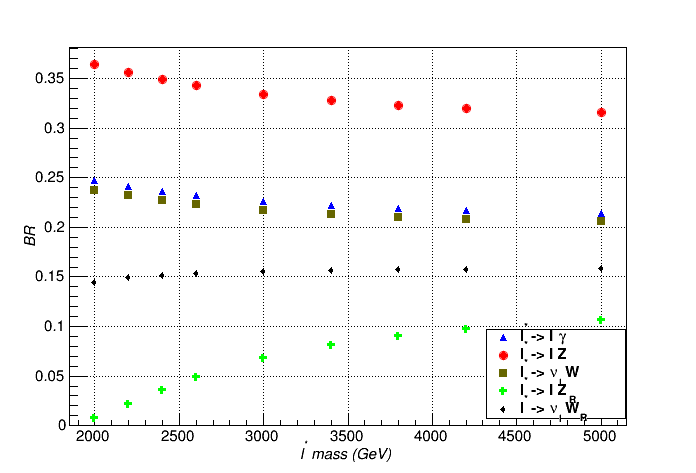}}
\end{tabular}
\caption{Branching Ratio as function of $\mu^*$ mass, $\Lambda = 5000$ GeV, $14$ TeV $pp$ beam.
  (a): BSZ, as in Ref. \cite{Baur:1990}
  (b-d): Left-right symmetric theory for $W^{'}$ mass 800 GeV and
         $Z'$ mass 800, 1200 and 1800 GeV, $f_1 = f_2 = f''$.}
\label{mubr}
\end{figure}

\begin{figure}
\begin{tabular}{c c}
\subfloat[]{\includegraphics[width=\figwidth]{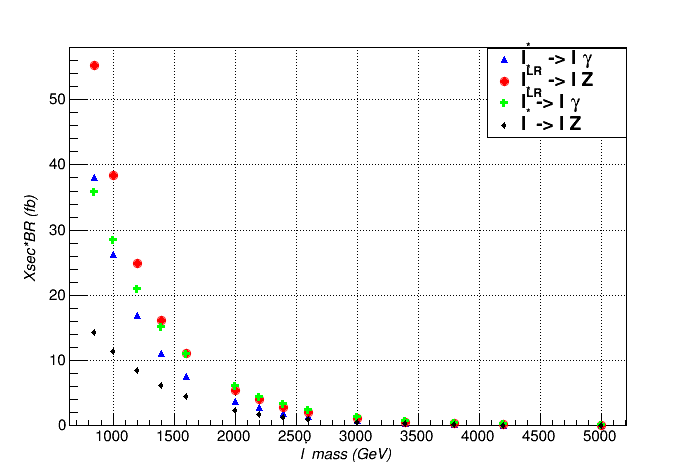}}
&
\subfloat[]{\includegraphics[width=\figwidth]{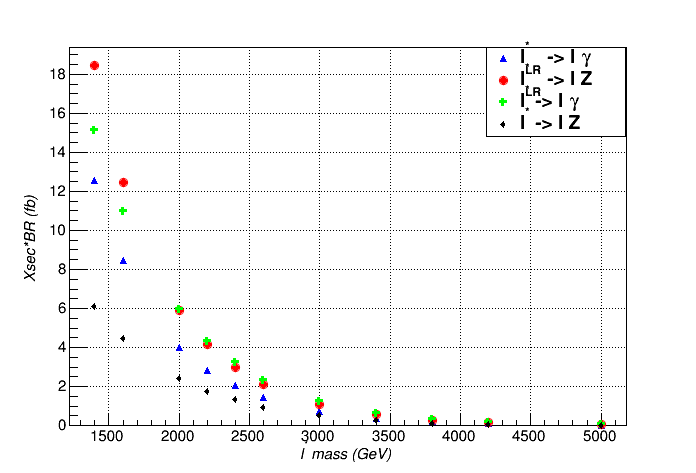}}
\\
\subfloat[]{\includegraphics[width=\figwidth]{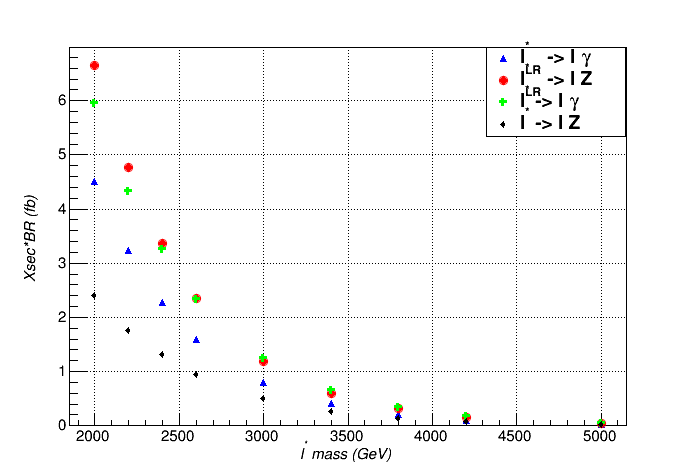}}
&
\subfloat[]{\includegraphics[width=\figwidth]{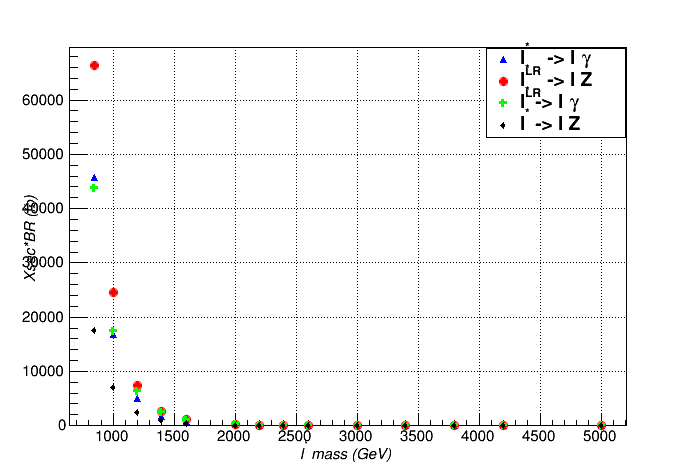}}
\\
\end{tabular}
\caption{
Comparison of cross section times branching ratios of $l^{*}$ for photon and $Z$ channels in 
the BSZ of Ref. \cite{Baur:1990} as well as
left-right symmetric theory, denoted by $l^{*}_{LR}$, as functions of $l^{*}$ mass, for $14$ TeV $pp$ beam, $f_1 = f_2 = f''$. 
For left-right symmetric theory, $W^{'}$ mass  is fixed to 800 GeV. 
(a)-(c): $Z'$ masses values 800, 1200, 1800 GeV, $\Lambda = 5000$ GeV.
(d): $Z'$ mass 800 GeV, $\Lambda = m_{\mu^*}$ GeV.
}
\label{muxsectimesbr}
\end{figure}

From the above plots we see that total production cross section, for
a given $m_{\mu^{*}}$, decreases
in the left right symmetric scenario as compared to the BSZ theory.
However the branching ratio of the $\mu^{*} \rightarrow \mu, Z$ channel
increases significantly, and in fact, the cross section times
branching ratio of this channel is much higher in the left right
symmetric theory. In fact, it is higher than even the cross section
times branching ratio of the $\mu^{*} \rightarrow \mu, \gamma$ channel for
all masses of $\mu^*$ and $Z'$.

The cross section times branching ratio of the $\mu^{*} \rightarrow \mu, \gamma$
channel is strongly depressed in the left-right symmetric theory as compared
to the BSZ theory. This effect is even more pronounced when 
$\Lambda = m_{\mu^*}$, which was the setting used by previous experiments
to set limits on $m_{\mu^*}$. 
Recently, the CMS experiment at the LHC has set the $e^{*}$ and $\mu^{*}$ mass limits
at $\sqrt{s} = 7 TeV$ with an integrated luminosity $L_{int} = 5 fb^{-1}$,
and excluded $m_{l^{*}} < 1.9$ TeV at 95\% C.L. \cite{CMS:2011}. 
Our work shows that if excited leptons were coupled to a left-right symmetric extension
of the SM, the excluded mass limit on excited leptons will come down significantly.
We therefore focus our study in the region $m_{l^{*}} > 850$ GeV. 

If excited leptons were coupled to a left-right symmetric extension
of the SM, the $l^{*} \rightarrow l Z$ channel, with its highest cross section
times branching ratio, is the best channel 
to be probed in a future LHC scenario of $14$ TeV beam energy.  
This channel with four lepton final states 
will have largest background contribution from SM direct production 
of ZZ pairs where both the Z bosons decay leptonically.
For this background, we get a LO cross section of $93 fb$ with Pythia8.
For an optimisitic scenario of $W'$ mass $800$ GeV, $Z'$ mass $800$ GeV and
signal  $l^{*} \rightarrow l Z \rightarrow l l' l'$ of about 5 fb,
a $5\sigma$-significance can be easily obtained with around $100 fb^{-1}$ of
data.

\section{Summary and Conclusions}
The LHC has searched for (left handed) excited electrons
and muons and ruled them out for masses below 1700 GeV. Their search
strategy was just to look for an excess in the $l^{*} \rightarrow l \gamma$
channel. If excited leptons were coupled to a left-right symmetric 
extension of the SM, our results
show that the cross section times branching ratio of this channel is
depressed by a factor of more than two. This implies that the excess
search carried out by Atlas would only rule out excited electrons and
muons having masses below 1000 GeV.

On the other hand, our results show that the cross section times branching
ratio of the $l^{*} \rightarrow l Z$ channel is enhanced in the left right symmetric
scenario, and in fact, is significantly more than that of the $l^{*}
\rightarrow l \gamma$
channel. This implies that the Z channel becomes a promising candidate
to search for left right symmetric excited leptons with masses above 1000 GeV.
This channel was never explored in earlier experimental searches. It should
be quite feasible to carry out this search with the current luminosity of
the LHC and the data already collected.

\begin{acknowledgments}

The work was supported by a Department of Science and Technology, India grant.
We thank Sudhir Vempati and Uma Sankar for helpful discussion.
\end{acknowledgments}

%\begin{bibliography}
\bibliographystyle{apsrev}
\bibliography{lstarLR_20141023}
%\end{bibliography}

\end{document}